\documentclass[final,3p,sort&compress,times,twocolumn,10pt]{elsarticle}
\usepackage[utf8]{inputenc}
\usepackage{amssymb}
\usepackage{graphicx,color}
\usepackage{relsize}
\usepackage{slashed}
\usepackage{ulem}
\usepackage{tabu}
\usepackage{booktabs}
\usepackage{bm}
\usepackage{amsmath}
\usepackage{amsthm}
\usepackage{mathrsfs}
\usepackage{fancyhdr}
\usepackage{flushend}
\usepackage{array}
\usepackage[all]{xy}
\usepackage{euscript}
\usepackage{enumerate}
\usepackage{slashed}
\usepackage{hyperref}
\usepackage{subfigure} 
\usepackage{epstopdf} 
\usepackage{xcolor}

\hypersetup{colorlinks=true,linkcolor=blue,citecolor=blue,menucolor=black,urlcolor=blue,filecolor=blue}

\journal{Physics Letters B}

\newcommand{\im}{{\rm Im}}
\newcommand{\re}{{\rm Re}}
\newcommand{\gev}{{\rm GeV}}
\newcommand{\mev}{{\rm MeV}}

\renewcommand{\vec}[1]{\mathbf{#1}}
\renewcommand{\arraystretch}{1.2}

\begin{document}

\begin{frontmatter}

\title{Extraction of $ND$ scattering lengths from the $\Lambda_b\rightarrow\pi^-pD^0$ decay and properties of the $\Sigma_c(2800)^+$}

\author[1]{Shuntaro Sakai}\ead{shsakai@itp.ac.cn}
\author[1,2]{Feng-Kun Guo}\ead{fkguo@itp.ac.cn}
\author[3]{Bastian Kubis}\ead{kubis@hiskp.uni-bonn.de}
\address[1]{CAS Key Laboratory of Theoretical Physics, Institute of Theoretical Physics, Chinese Academy of Sciences, Beijing 100190, China}
\address[2]{School of Physical Sciences, University of Chinese Academy of Sciences, Beijing 100049, China}
\address[3]{Helmholtz-Institut f\"ur Strahlen- und Kernphysik (Theorie) and Bethe Center for Theoretical Physics, Universit\"at Bonn, 53115 Bonn, Germany}

\begin{abstract}
The isovector and isoscalar $ND$
$s$-wave scattering lengths are extracted by fitting to the LHCb data of the $pD^0$ invariant-mass distribution
in the decay $\Lambda_b\rightarrow\pi^-pD^0$, making use of the cusp effect at the $nD^+$ threshold.
The analysis is based on a coupled-channel nonrelativistic effective field theory.
We find that the real part of the isovector $ND$ scattering length is unnaturally large due to the existence of a 
near-threshold state with a mass around 2.8~GeV.
The state is consistent with the $\Sigma_c(2800)^+$ resonance observed at Belle. 
Our results suggest that it couples strongly to the $ND$ channel in an $s$-wave, and that its quantum numbers are $J^P=1/2^-$. The strong cusp behavior at the $nD^+$ threshold can be verified using updated LHCb data.
\end{abstract}


\end{frontmatter}

\section{Introduction}
\label{sec:intro}

Properties of the $ND$ system have been paid much attention to
as an analogue to the $\bar{K}N$ system,
where there exists the $\Lambda(1405)$ resonance as a $\bar{K}N$ quasi-bound state near threshold~\cite{Dalitz:1960du,Kaiser:1995eg,Oset:1997it,Oller:2000fj}
(for reviews, see Refs.~\cite{Hyodo:2011ur,Kamiya:2016jqc} and the review article dedicated to the $\Lambda(1405)$ in the Reviews of Particle Physics~\cite{Tanabashi:2018oca}).
Several studies are devoted to the $ND$ system 
from the viewpoint of the description of the $\Lambda_c^*$ or $\Sigma_c^*$ charmed baryons~\cite{Lutz:2003jw,Hofmann:2005sw,Lutz:2005vx,Mizutani:2006vq,GarciaRecio:2008dp,JimenezTejero:2009vq,Romanets:2012hm,Haidenbauer:2010ch,Liang:2014kra,Carames:2014mva,Garcia-Recio:2015jsa,Hofmann:2006qx,Dong:2010gu,Zhang:2012jk,Wang:2020dhf}.
Some reactions concerning the $ND$ system and $\Lambda_c^*$ resonances have been investigated~\cite{Haidenbauer:2008ff,Liang:2016ydj,Liang:2016exm} 
and extensions to systems with $D$ and nuclei performed 
(see, for example, Ref.~\cite{Tolos:2013gta} and references therein). Most of these studies discuss possible hadronic molecules (see Ref.~\cite{Guo:2017jvc} for a recent review) in the charmed-meson--nucleon systems, analogous to the $XYZ$ states in the heavy-quarkonia mass region and the hidden-charm pentaquarks.
However, relevant experimental information is scarce and the above-mentioned calculations are based on phenomenological models.

At low energies, the short-range interaction between two particles can be described by the effective-range expansion,\footnote{The sign convention here is such that a positive (negative) scattering length corresponds to an attractive (repulsive) interaction in the absence of a bound-state pole. When there is a bound state below threshold, the scattering length is negative.} 
\begin{equation}
 f_0^{-1}(k) = \frac1{a_0} + \frac12 r_{0}k^2 - i\,k + \mathcal{O}(k^4)\label{eq:ere}
\end{equation}
for $s$-waves, where $f_0(k)$ is the $s$-wave amplitude and $k$ is the magnitude of the center-of-mass (c.m.) momentum. Threshold parameters, including the scattering length $a_0$ and the effective range $r_{0}$, are important quantities as they govern the low-energy behavior of the scattering amplitude. Their values can be used to infer the structure of $s$-wave shallow bound states if there are any~\cite{Weinberg:1965zz,Guo:2017jvc}.

It is well-known that in invariant-mass distributions there must be cusps exactly at two-body $s$-wave thresholds of opening channels due to unitarity. Because the masses of the involved particles are fixed, the cusp strength is then determined by the interaction at threshold, and thus threshold cusps can be used to extract the corresponding scattering lengths (we refer to Ref.~\cite{Guo:2019twa} for a recent review on this topic).
In the case of $\pi\pi$ scattering, the role of a cusp at the $\pi^+\pi^-$ threshold in the $\pi^0\pi^0$ spectrum was discussed  in Refs.~\cite{Budini:1961bac,Meissner:1997fa},
and a possible way to extract the $\pi\pi$ scattering length from the threshold cusp was suggested in Refs.~\cite{Cabibbo:2004gq,Cabibbo:2005ez}, 
followed by reformulations based on a nonrelativistic effective field theory (NREFT)~\cite{Colangelo:2006va,Gasser:2011ju} (cf.\ also Ref.~\cite{Gamiz:2006km}). The difference between the $\pi\pi(I=0)$ and $\pi\pi(I=2)$ scattering lengths is tied to the magnitude of
the $\pi^+\pi^-$ threshold cusp in the $\pi^0\pi^0$ distribution in the decays $K^\pm\rightarrow\pi^\pm\pi^0\pi^0$,
and the $\pi\pi$ scattering lengths were determined from the cusp in the subsequent experimental studies~\cite{Batley:2005ax,Batley:2000zz}.
This framework was applied to other processes in Refs.~\cite{Bissegger:2007yq,Gullstrom:2008sy,Kubis:2009sb,Hyodo:2011js,Liu:2012dv}.
In the present study, we try to extract the $s$-wave scattering lengths of the $ND$ system from the $pD^0$ invariant-mass distribution of the decay $\Lambda_b\rightarrow\pi^-pD^0$ measured by the LHCb Collaboration~\cite{Aaij:2017vbw}, where a peculiar structure near the $nD^+$ threshold (about 6~MeV higher than the $pD^0$ threshold) is seen. 

The Letter is organized as follows. In Sec.~\ref{sec:setup}, the NREFT formalism is introduced that can be used in analyzing the $pD^0$ invariant-mass distribution in the near-threshold region. In Sec.~\ref{sec:results}, the results of fitting to the $s$-wave contribution extracted from the LHCb analysis are presented. A brief summary is given in Sec.~\ref{sec:summary}.

\section{Setup}
\label{sec:setup}

\begin{figure}[t]
 \centering
 \includegraphics[width=6.6cm]{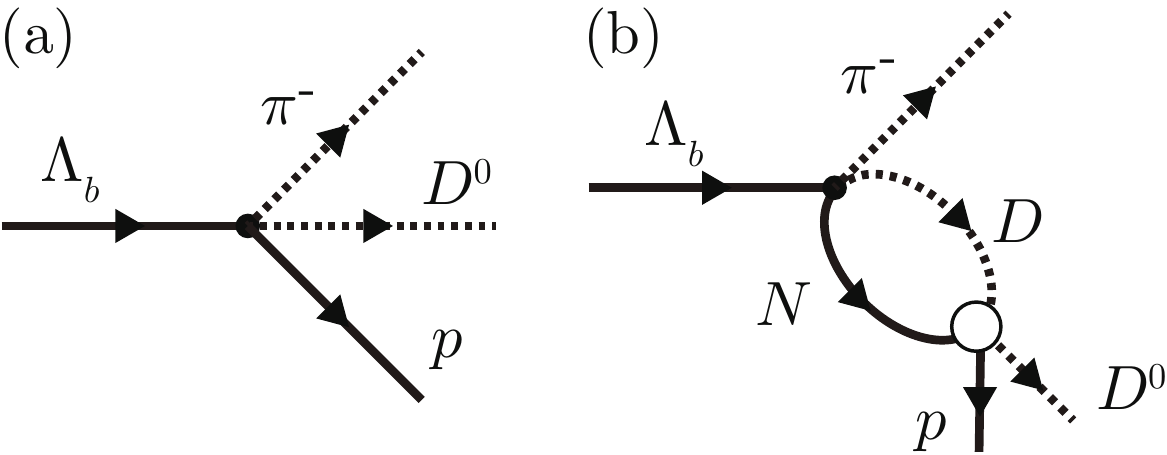}
 \caption{Diagrams for the decay $\Lambda_b\rightarrow\pi^-pD^0$ taken into account in this study. Here $ND$ represents both $pD^0$ and $nD^+$.}
 \label{fig:diag1}
\end{figure}

In Fig.~\ref{fig:diag1}, we show the diagrams considered in this study.
Diagram (a) of Fig.~\ref{fig:diag1} represents the $\pi^-pD^0$ production from $\Lambda_b$ without rescattering of particles,
and diagram (b) takes into account the subsequent rescattering of $ND\to pD^0$ where $ND$ represents both $pD^0$ and $nD^+$.

First, we explain the $ND$ scattering part. The $\pi\pi$ low-energy scattering is known to be rather weak due to chiral suppression; on the contrary, there is no suppression for the $ND$ near-threshold interaction and it might be strong enough to generate a nearby pole.
We employ the coupled-channel NREFT as developed in Refs.~\cite{Cohen:2004kf,Braaten:2005jj}, which is based on the Lippmann--Schwinger equation and is adequate to treat the $ND$ scattering in the near-threshold region. There are two channels: $pD^0$ and $nD^+$.
The nonrelativistic $T$-matrix
is given by the Lippmann--Schwinger equation as~\cite{Cohen:2004kf,Braaten:2005jj}
\begin{align}
 t_{ij}&= \left[(1-v\,G)^{-1}v\right]_{ij} \nonumber\\ 
 &=\frac{2\pi}{\det}
 \begin{pmatrix}
  \frac{1}{\mu_{1}}\left(-\frac{1}{a_{22}}+ip_{2}\right)&-\frac{1}{a_{12}\sqrt{\mu_{1}\mu_{2}}}\\
  -\frac{1}{a_{12}\sqrt{\mu_{1}\mu_{2}}}&\frac{1}{\mu_{2}}\left(-\frac{1}{a_{11}}+ip_{1}\right) 
 \end{pmatrix}_{ij},
 \label{eq:tND}\\
 \det&= \left(\frac{1}{a_{12}}\right)^2 - \left(\frac{1}{a_{11}}-ip_{1}\right)\left(\frac{1}{a_{22}}-ip_{2}\right), 
 \label{eq:det}
\end{align}
where $i=1$ and $2$ denote the $pD^0$ and $nD^+$ channels, respectively, $\mu_i$ denotes the reduced mass,
and $p_i$ is the nonrelativistic momentum of $D$ (or $N$) in the $ND$ c.m. frame, 
\begin{align}
 p_i=\sqrt{2\mu_i(M_{ND}-M_i-m_i)}\,, 
\end{align}
with $M_{ND}$ being the $ND$ invariant mass, and $M_i$ $(m_i)$ the mass of $D^0$ or $D^+$ ($p$ or $n$). With the above expression, the momentum $p_i$ is defined above threshold. Below threshold, it is analytically continued to  $i\sqrt{2\mu_i(M_i+m_i-M_{ND})}$. 
The NREFT is based on an expansion in powers of the velocity of the particles in their c.m.\ frame. At the leading order 
of the NREFT, which is sufficient in the immediate vicinity of the thresholds given the current data quality, the spins of the involved particles do not need to be considered. The interaction kernel $v$ is a matrix for constant contact terms (the next-to-leading order terms are of $\mathcal{O}(p_i^2/\mu_i^2)$), and $G$ is a $2\times2$ diagonal matrix with the diagonal matrix element
$G_i$ given by the nonrelativistic two-body loop function,
\begin{align}
 G_{i}^\Lambda &= \int^\Lambda\frac{d^3\vec q}{(2\pi)^3} \frac{2\mu_i}{\vec q^2 - p_i^2 - i\epsilon} \nonumber\\
 &= i\frac{\mu_i}{2\pi}p_i +  \frac{\mu_i}{\pi^2}\Lambda \left[ 1+ \mathcal{O}\left( \frac{p_i^2}{\Lambda^2} \right) \right].
 \label{eq:G}
\end{align}
The ultraviolet (UV) divergence in the loop integral is regularized using a three-momentum cutoff $\Lambda$.
In Eq.~\eqref{eq:tND}, 
the cutoff dependence of the loop function $G_i$ is absorbed by the interaction kernel $v$, and the resulting parameters $a_{ij}$ are cutoff-independent
(see Refs.~\cite{Cohen:2004kf,Braaten:2005jj} for details).
The $ND$ scattering lengths can be expressed in terms of these parameters.
The scattering length $a_i$ in channel $i$ is defined by the scattering amplitude at threshold as
\begin{align}
 a_i  = \frac{\mu_i}{2\pi}t_{ii}\bigg|_{M_{ND}=m_i+M_i}.\label{eq:sl_def}
\end{align}
If the interaction is strong enough, a near-threshold pole of the $T$-matrix can be generated as a zero of the determinant defined in Eq.~\eqref{eq:det}.

With the isospin phase convention
\begin{equation}
 \left|D^+\right\rangle=-\left|I=\frac12,I_z=\frac12\right\rangle
\end{equation}
and all other states taking a positive sign, the $ND$ scattering amplitudes in the isospin basis and that in the particle basis 
are related to each other as
\begin{align}
 t_{ND(I=0)}=&\frac{1}{2}\left(t_{pD^0,pD^0}+t_{nD^+,nD^+}+2t_{nD^+,pD^0}\right), \nonumber\\ 
 t_{ND(I=1)}=&\frac{1}{2}\left(t_{pD^0,pD^0}+t_{nD^+,nD^+}-2t_{nD^+,pD^0}\right). \label{eq:tI} 
\end{align}
Hence, the difference of the $ND$ scattering lengths with $I=0$ and $I=1$ is proportional to $t_{nD^+,pD^0}$,
\begin{equation}
    t_{nD^+,pD^0} = \frac12 \left(t_{ND(I=0)} - t_{ND(I=1)} \right) .
\end{equation}

At the lower ($pD^0$) threshold, the $T$-matrix elements are
\begin{align}
  t_{11}^\text{th} &= \frac{2\pi}{\mu_1}\left(-\frac{1}{a_{22}}-\kappa\right) \left[\frac1{a_{12}^2} - \frac1{a_{11}} \left(\frac1{a_{22}} + \kappa \right)\right]^{-1} \nonumber\\
  &\equiv  \frac{2\pi}{\mu_1}a_{c} \simeq  \frac{2\pi}{\mu_1} \left[ \frac1{a_{11}} - \frac1{a_{12}^2 \left(a_{11}^{-1} +\kappa \right) } \right]^{-1} , \nonumber\\
  t_{12}^\text{th} &= -\frac{2\pi}{\sqrt{\mu_1\mu_2}} \frac1{a_{12}} 
    \left[ \frac1{a_{12}^2} - \frac1{a_{11}} \left(\frac1{a_{22}} + \kappa \right) \right]^{-1} \nonumber\\
    &\equiv  \frac{2\pi}{\sqrt{\mu_1\mu_2}}a_x \simeq -\frac{2\pi}{\sqrt{\mu_1\mu_2}} \frac{a_{11}}{a_{12}} \left(\frac{a_{11}}{a_{12}^2} -\frac1{a_{11}} -\kappa \right)^{-1} , 
    \label{eq:tij}
\end{align}
where  $\kappa=\sqrt{2\mu_2\delta}$ with $\delta = m_n+M_{D^+}-m_p-M_{D^0}$. One sees that $\kappa = \mathcal{O}\left({\delta}^{1/2}\right)$. The parameters $a_{ij}$ must be analytic in the involved hadron masses. Thus, the isospin breaking difference $a_{22}-a_{11}$ is of $\mathcal{O}(\delta)$, and we have neglected such a difference in Eq.~\eqref{eq:tij}.
The more natural parameters are $a_c$ and $a_{x}$ instead of $a_{11}$ and $a_{12}$.
They are connected to the isoscalar and isovector scattering lengths as 
\begin{align}
  a_c &= \frac12\left(a_{ND(I=0)} + a_{ND(I=1)}\right), \nonumber\\
  a_x &= \frac12\left(a_{ND(I=0)} - a_{ND(I=1)}\right).
  \label{eq:acax}
\end{align}
The parameters $a_{11}$ and $a_{12}$ can be expressed in terms of  $a_c$ and $a_{x}$:
\begin{align}
 a_{11} = \frac{ a_{c}^2 -a_x^2 }{a_{c}+a_x^2 \kappa}, \quad
 a_{12} = \frac{ a_{c}^2 -a_x^2 }{a_x(1+ a_{c}\kappa)} .
\end{align}

\begin{figure}[t]
 \centering
 \includegraphics[width=5cm]{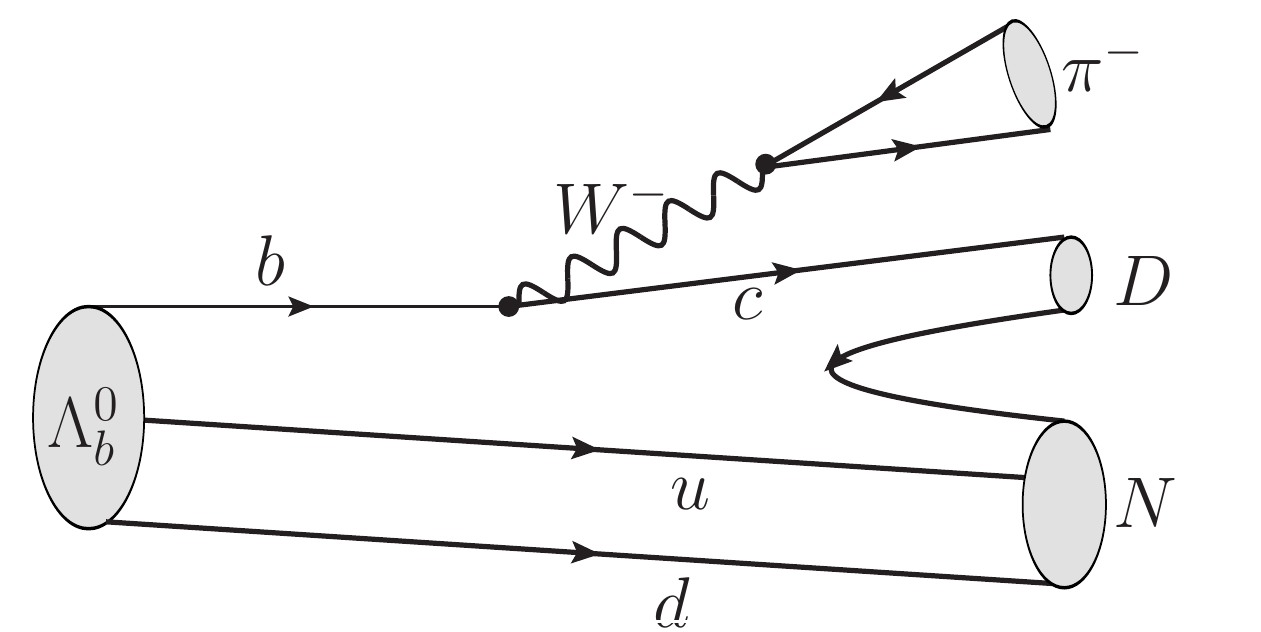}
 \caption{The $W^-$ emission mechanism for the $\Lambda_b\rightarrow\pi^-pD^0$ decay.}
 \label{fig:diag2}
\end{figure}
Second, let us consider the production from the $\Lambda_b$ decays. The leading contribution to the $\Lambda_b \to \pi^- ND$ decay in terms of the Cabibbo--Kobayashi--Maskawa matrix and color counting is the $W^-$ emission mechanism, where the $W^-$ boson emitted from the $b\to c W^-$ process becomes a $\pi^-$, see Fig.~\ref{fig:diag2}. Because the $ud$ pair in the $\Lambda_b$ is an isoscalar, the produced $ND$ pair would be an isospin singlet as well. Thus, the production of the $pD^0$ should be approximately the same as that of the $n D^+$. 

Since parity is not conserved in the $\Lambda_b\to\pi^-ND$ weak decay, the $ND$ pair can be produced not only in the $s$-wave, but also in higher partial waves. With the statistics of the current data, it is not possible to fix parameters with different partial waves included. 
Thus, we take the LHCb data and subtract the contributions from partial waves other than the $ND$ $s$-wave. Several fits are presented in the LHCb analysis~\cite{Aaij:2017vbw}. We take the analysis presented in Fig.~12 of Ref.~\cite{Aaij:2017vbw}, and subtract the contributions from the $J^P=1/2^+$ and $3/2^\pm$ partial waves from the measured $pD^0$ invariant-mass distribution. In this way, only the $1/2^-$ part, corresponding to the $ND$ $s$-wave, is left.

For the energy region close to the thresholds, the $s$-wave production can be approximated by a constant contact term followed by the final-state interaction (FSI). The $ND$ FSI can be described by the nonrelativistic $T$-matrix discussed above. Any possible singular behavior comes from the rescattering given in Eq.~\eqref{eq:tND}. 
We parametrize the $s$-wave contact term for the $\Lambda_b\rightarrow\pi^-ND$ production amplitude by a constant $V_P$.
A momentum factor $p_{\pi^-}$ associated with the $W^-$ transition to $\pi^-$ is absorbed into $V_P$ as a constant
because this factor is irrelevant to the $nD^+$ threshold cusp of interest, and we are focusing on a very small range of the $ND$ invariant mass around the $pD^0$ and $nD^+$ thresholds.

Taking into account the $ND$ rescattering described by Eq.~\eqref{eq:tND} and the fact that $ND$ produced from the $\Lambda_b$ weak decay has $I=0$,
the decay amplitude of $\Lambda_b\rightarrow\pi^-pD^0$ in Fig.~\ref{fig:diag1} for an $ND$ $s$-wave can be written as follows:
\begin{align}
 \mathcal{A}_{s\text{-wave}}=&V_P^\Lambda \left(1 
 +G_{pD^0}^\Lambda t_{pD^0,pD^0}+G_{nD^+}^\Lambda t_{nD^+,pD^0}\right),\label{eq:decay_amp}
\end{align}
where $V_P$ has been rewritten as $V_P^\Lambda$ to emphasize that it depends on the cutoff $\Lambda$.
The second and third terms in the parentheses come from diagram~(b) in Fig.~\ref{fig:diag1} with $s$-wave $ND$ rescattering.
The rescattering of the other pairs ($\pi N$ and $\pi D$) does not matter either because the pion moves much faster than the $D$-meson and the nucleon in the near-$ND$-threshold region.
One notices that the $T$-matrix elements $t_{pD^0,pD^0}$ and $t_{nD^+,pD^0}$ are physical quantities and do not depend on the cutoff introduced in regularizing the UV divergence. 
The cutoff dependence of the loop functions in Eq.~\eqref{eq:decay_amp} needs to be absorbed by the production vertex $V_P^\Lambda$. Such a requirement is fulfilled by a multiplicative renormalization with $V_P\propto 1/\Lambda$ and by keeping only the leading-order term (in the expansion in powers of $p_i/\Lambda$) in the loop function in Eq.~\eqref{eq:G}. 
Therefore, we find the following cutoff-independent amplitude:
\begin{align}
 \mathcal{A}_{s\text{-wave}}=V_P \left(t_{pD^0,pD^0} + t_{nD^+,pD^0}\right),
 \label{eq:decay_amp_r}
\end{align}
where $V_P$ has been redefined to absorb the cutoff as well as other multiplicative constants with approximating $\mu_{pD^0}\simeq \mu_{nD^+}$ in the cutoff term of $G_i^\Lambda$.
The procedure can be understood as that the process is dominated by the contribution from diagram (b) in Fig.~\ref{fig:diag1}.

Consequently, the $1/2^-$ part of the $pD^0$ invariant-mass distribution for the decay $\Lambda_b\rightarrow\pi^-pD^0$ can be fitted using 
\begin{align}
 \frac{d\Gamma_{\Lambda_b\rightarrow\pi^-pD^0}}{dM_{pD^0}}= \mathcal{N} p_{\pi^-}p_{D^0}\left|t_{pD^0,pD^0} + t_{nD^+,pD^0}\right|^2\,,
\label{eq:md}
\end{align}
where $\mathcal{N}$ is an overall normalization constant, $p_{\pi^-}$ ($p_{D^0}$) is the momentum of the $\pi^-$ ($D^0$) in the $\Lambda_b$ $(pD^0)$ rest frame.

The $ND$ scattering lengths can be extracted by fitting to the $1/2^-$ partial wave of the $pD^0$ invariant-mass distribution reported in Ref.~\cite{Aaij:2017vbw} using Eq.~\eqref{eq:md} as a fitting function. 
There is one more complexity due to the inelasticity from  coupling $p D^0$ and $nD^+$ to lower channels such as $\pi\Lambda_c$ and $\pi\Sigma_c$. Since these channels are far away from the energy region of interest, the inelastic effects may be included by introducing imaginary parts to the parameters $a_x$ and $a_c$. 
Unitarity is useful to constrain the parameter space: $\im\,t_{ii} = \sum_j \rho_j |t_{ij}|^2\geq0$ where $\rho_j\geq0$ is the phase space factor of the intermediate states $j$.
In order to obtain a stable fit, we further reduce the number of free parameters by approximating 
\begin{equation}
  \im\,a_x \simeq-\im\,a_c.
  \label{eq:imapp}
\end{equation}
Considering that the phase space for the lowest two-body isovector channel $\pi\Lambda_c$ is larger than that for the isoscalar channel $\pi\Sigma_c$, it is reasonable to assume $\im\,a_{ND(I=0)}\ll \im\,a_{ND(I=1)}$. We therefore treat $\im\,a_{ND(I=0)}$ as a higher-order effect and neglect it in the leading approximation. 
This approximation is supported by the existing model calculations, listed in Table~\ref{tab:ref-scat} below.
Then Eq.~\eqref{eq:imapp} follows from Eq.~\eqref{eq:acax}.
Hence, there are four free parameters in the fit, including $\re\,a_c$, $\im\,a_c$, $\re\,a_x$, and $\mathcal{N}$.

\section{Results}
\label{sec:results}

\begin{figure}[t]
 \centering
 \includegraphics[width=\linewidth]{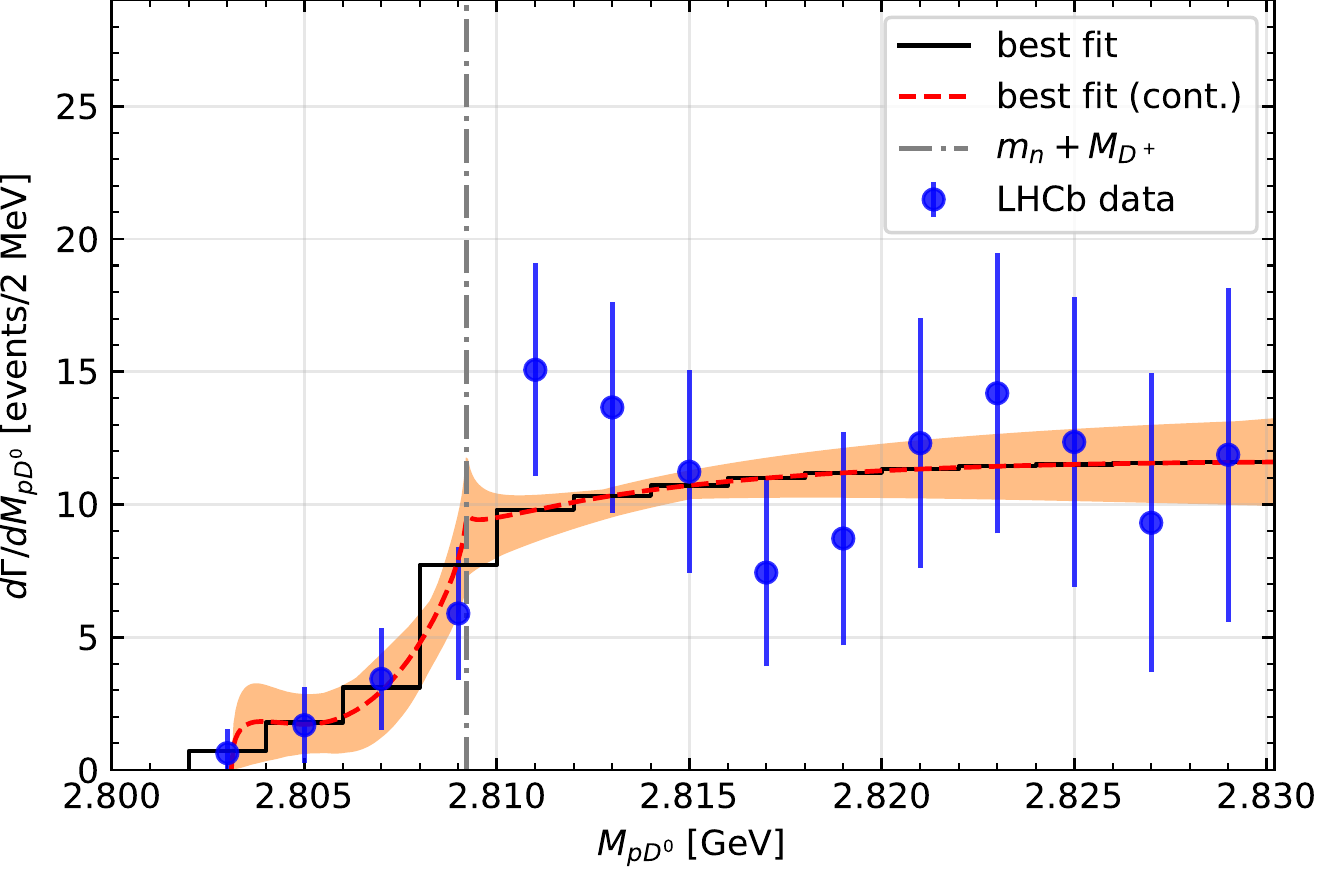}
 \caption{The best fit to the $pD^0$ distribution.
 The histogram shows the best fit with event numbers integrated in each bin, the dashed curve is the corresponding continuous distribution, which exhibit a clear cusp at the $nD^+$ threshold, and the band is the corresponding $1\sigma$ error region.
 The data points with error bars are taken from Fig.~12(b) in Ref.~\cite{Aaij:2017vbw} with the contributions from the $J^P=1/2^+$ and $3/2^\pm$ partial waves subtracted.
 The vertical dot-dashed line denotes the $nD^+$ threshold.
 }
 \label{fig:res1}
\end{figure} 

\begin{figure}[t]
 \centering
 \includegraphics[width=\linewidth]{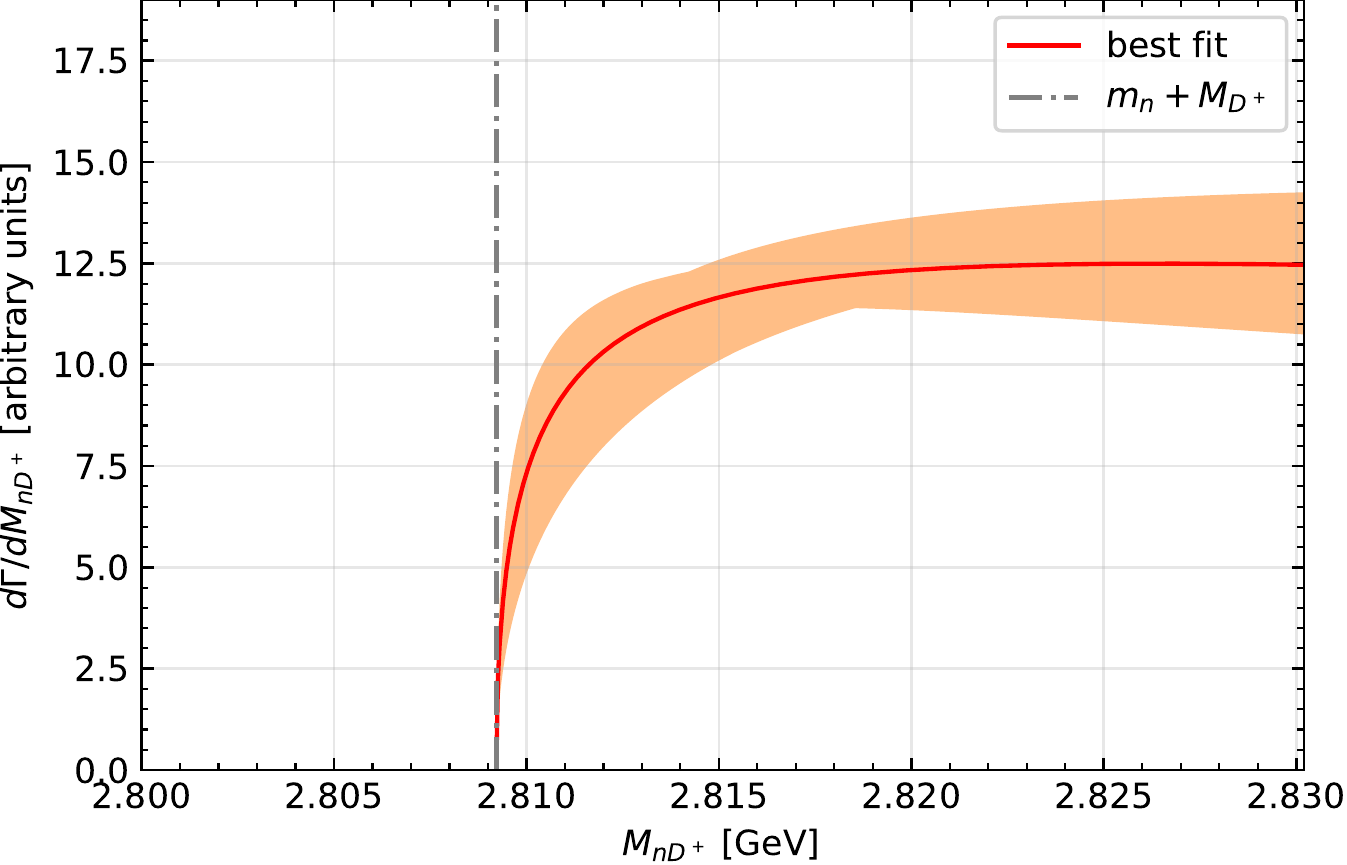}
 \caption{Prediction of the $nD^+$ distribution for the decay $\Lambda_b\to \pi^-nD^+$. The vertical dot-dashed line denotes the $nD^+$ threshold.
 } 
 \label{fig:prediction}
\end{figure} 

\begin{table}[tb]
\renewcommand{\arraystretch}{1.2}
 \centering
 \caption{Parameters from the best fit. The uncertainties are the $1\sigma$ errors propagated from the statistical uncertainties of the data.  }
 \medskip
 \begin{tabular}{cccc}
  \toprule
   $\re\,a_{c}$  &$\im\,a_{c}$ &$\re\,a_{x}$ &$\mathcal{N}$ \\
   $(\gev^{-1})$ &$(\gev^{-1})$ &$(\gev^{-1})$ &$(\gev^{2})$ \\ \midrule
   $-11.7^{+4.5}_{-5.8}$ & $6.8^{+4.0}_{-6.8}$ & $7.7^{+4.7}_{-3.4}$ & $0.04^{+0.06}_{-0.02}$ \\
  \bottomrule
 \end{tabular}
 \label{tab:res1}
 \renewcommand{\arraystretch}{1.0}
\end{table}

\begin{table*}[t]
\renewcommand{\arraystretch}{1.2}
 \centering
 \caption{The $DN$ scattering lengths obtained from our fit in comparison with the results calculated in selected  phenomenological models.
 The values for the SU(4) $DN$ model~\cite{Mizutani:2006vq} and the SU(8) $DN$ model~\cite{GarciaRecio:2008dp} are taken from Table~2 of Ref.~\cite{Haidenbauer:2010ch}, which considers a meson-exchange model. All values are given in units of fm.}
 \begin{tabular}[t]{l| ccccc}
  \toprule
  $a_{ND}$ [fm] & Our result  &SU(4)~\cite{Lutz:2005vx} &SU(4)~\cite{Mizutani:2006vq} &SU(8)~\cite{GarciaRecio:2008dp} &Meson-exchange model~\cite{Haidenbauer:2010ch} \\\midrule
  ${I=0}$ & $-0.79^{+0.66}_{-0.61}$ 
  &$-0.43$ &$-0.57+i\,0.001$ &$0.004+i\,0.002$ &$-0.41+i\,0.04$\\
  ${I=1}$ & $-3.8^{+1.4}_{-2.0} + i\, 2.7^{+1.6}_{-2.7}$ 
  & $-0.41$ &$-1.47+i\,0.65$ &$0.33+i\,0.05$ &$-2.07+i\,0.57$\\
  \bottomrule
 \end{tabular}
 \label{tab:ref-scat}
 \renewcommand{\arraystretch}{1.0}
\end{table*}

As mentioned above, we subtract the $1/2^+$ and $3/2^\pm$ contributions from the $pD^0$ invariant-mass distribution of the $\Lambda_b\rightarrow \pi^-pD^0$ decay as reported in Fig.~12(b) in Ref.~\cite{Aaij:2017vbw}.
Using the MINUIT algorithm~\cite{James:1975dr,iminuit}, we fit to the 14 data points below 2.83~GeV by averaging the  function in Eq.~\eqref{eq:md} for each bin to take into account the binning of the measured $pD^0$ invariant-mass distribution.\footnote{The hard momentum scale of the NREFT considered here can be estimated as $\Lambda_\text{hard} = \sqrt{2\mu_i (M_{D^*}-M_D)}\simeq 0.42$~GeV, which is the scale for the opening of the next relevant threshold, $ND^*$. At $M_{ND}=2.83$~GeV, $p_i^2/\mu_i^2$ is smaller than 0.09, meaning that the nonrelativistic expansion is well justified, and $p_{i}^2/\Lambda_\text{hard}^2$ is 0.19 for $pD^0$ and 0.15 for $nD^+$, which is an estimate of the relative importance of the next-to-leading order contribution.} Thus, the leading-order NREFT treatment is sufficient, given the large uncertainties of the current data set. In this range, the nonrelativistic treatment of the $ND$ systems is well justified. 
The best fit has a reduced chi-square of $\chi^2/\text{dof}=4.9/10$. 
In Fig.~\ref{fig:res1}, the best fit to the $1/2^-$ partial wave of the $pD^0$ distribution is compared to the data.
The measured distribution indeed shows an evident change around the $nD^+$ threshold despite the low statistics, and the best fit curve has a clear $nD^+$ threshold cusp.
The resulting parameters from the fit are given in Table~\ref{tab:res1}. We have checked that the values of $a_c$ and $a_x$ remain almost the same if we keep the loop functions in Eq.~\eqref{eq:decay_amp} and use $\Lambda$ ranging from 0 to 10~GeV (the expression of $G^\Lambda$ with $\Lambda=0$~GeV is formally the same as the one using the $\overline{\rm MS}$ scheme of dimensional regularization as one does in the studies of the $\pi\pi$ cusp in NREFT, see, e.g., Refs.~\cite{Colangelo:2006va,Gasser:2011ju}). The change of $\Lambda$ is absorbed by a corresponding change in the value of the normalization ($V_P^2$ effectively).
Using the parameters from the fit, the $nD^+$ distribution for the decay $\Lambda_b\to \pi^-nD^+$ can be easily predicted, which is shown in Fig.~\ref{fig:prediction}.

The isoscalar and isovector $ND$ scattering lengths obtained from the fit follow from Eq.~\eqref{eq:acax}, and are listed in Table~\ref{tab:ref-scat}, where the errors are the $1\sigma$ uncertainties propagated from the statistical errors of the data.
For comparison, we also show the scattering lengths obtained in a few phenomenological models~\cite{Lutz:2005vx,Mizutani:2006vq,GarciaRecio:2008dp,Haidenbauer:2010ch} in Table~\ref{tab:ref-scat}. One sees that the results of the SU(4) $DN$ model~\cite{Mizutani:2006vq} and the meson-exchange model~\cite{Haidenbauer:2010ch} are compatible with our findings.
\footnote{The $nD^0$ scattering length is evaluated to be $-0.764+i\,0.615$~fm in Ref.~\cite{Raha:2017ahu} (here the convention of the scattering length has been changed to be the same as ours), in which the parameters are chosen to reproduce the mass and width of the neutral $\Sigma_c(2800)$ measured by Belle~\cite{Mizuk:2004yu}.}

\begin{figure}[t]
 \centering
 \includegraphics[width=\linewidth]{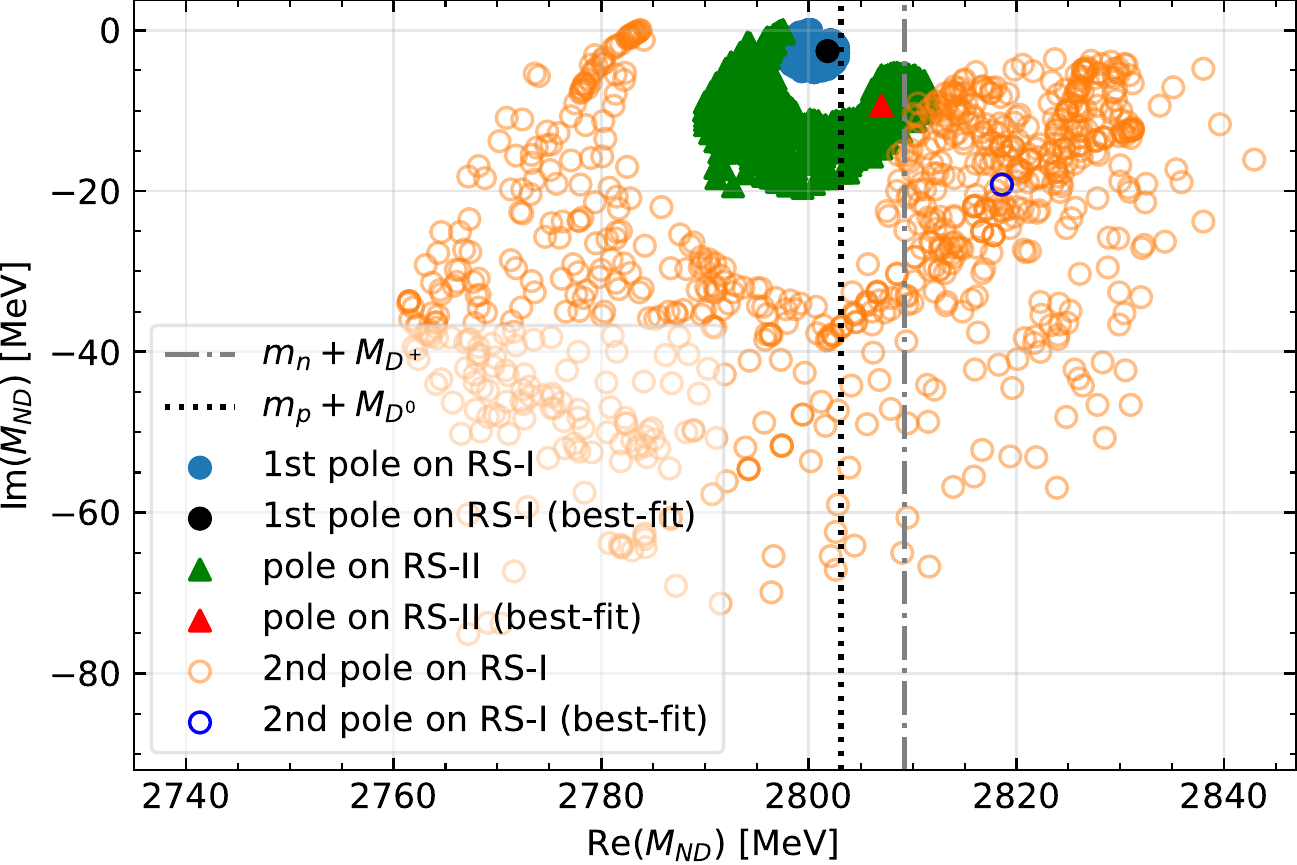}
 \caption{Poles in the first and second Riemann sheets using parameters in the $1\sigma$ range of the best fit.}
 \label{fig:poles}
\end{figure}

It is worthwhile to notice that the real parts of both the isoscalar and isovector scattering lengths are negative, meaning that the interaction in each channel is either repulsive or strongly attractive such that there is a bound state below threshold.
The large absolute value of $\re\,a_{ND(I=1)}$ within errors suggests the existence of a near-threshold isovector bound state; in contrast, no strong conclusion can be made in the isoscalar sector: within the uncertainties, one can find a pole, which can be near threshold or too far away to be valid in the NREFT framework.
Indeed, keeping isospin symmetry breaking, we find two poles in the first Riemann sheet (RS-I) of the complex $M_{ND}$ plane\footnote{The first Riemann sheet is defined as the Riemann sheet with Im$\,p_{1}>0$ and  Im$\,p_{2}>0$, and second Riemann sheet is defined as the one with  Im$\,p_{1}<0$ and  Im$\,p_{2}>0$. It is worthwhile to notice that because of the complexity of the $a_{11,12}$ parameters from the inelastic channels, there can be poles off the real axis on the first Riemann sheet, and the Schwarz reflection principle is not respected.} for the $T$-matrix defined in Eq.~\eqref{eq:det}, and one pole on the second Riemann sheet (RS-II). Their locations computed using about 1000 parameter sets within $1\sigma$ are shown in Fig.~\ref{fig:poles}. 
\begin{figure}[t]
 \centering
 \includegraphics[width=\linewidth]{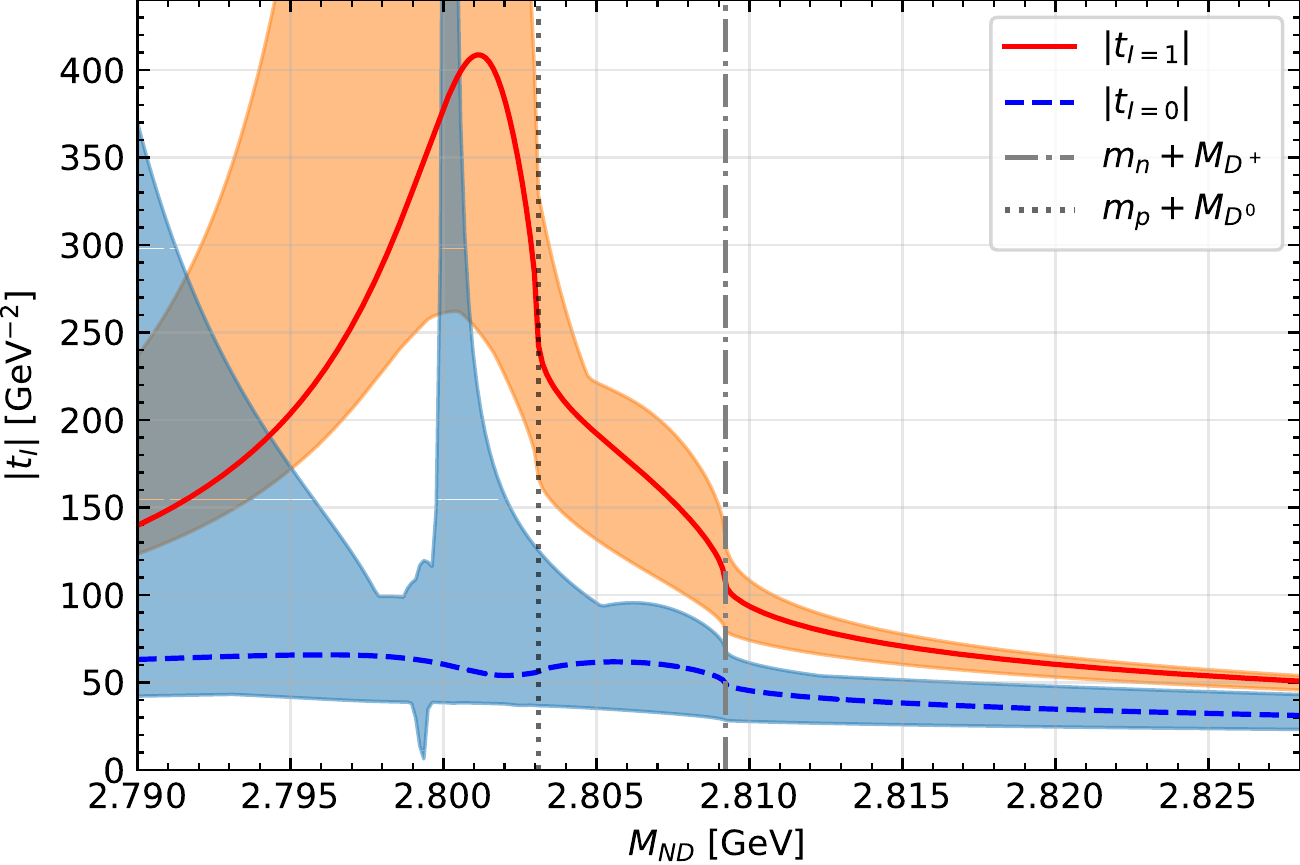}
 \caption{The absolute values of the isovector and isoscalar $T$-matrix elements from the best fit. The band is the $1\sigma$ uncertainty propagated from the data.}
 \label{fig:abst}
\end{figure}
For the first pole on RS-I, the absolute value of its residue to the isovector channel is at least more than 6 times (up to two orders of magnitude) larger than that to the isoscalar channel, except when the pole is located on the real axis, in which case the residue sizes are comparable. The pole on RS-II couples more strongly to the isovector than to the isoscalar channel, with the ratio of the residue size ranging from 1.6 to 3.8.\footnote{In the isospin symmetric limit, there is only one isovector pole.} These two poles are at:
\begin{align}
  \text{RS-I:~}&~~ 2801.8^{+1.0}_{-4.0} - i\,(2.6\pm2.6)~\mev, \nonumber\\
  \text{RS-II:~}&~~ 2807.0^{+4.5}_{-17.1} - i\left(9.4^{+10.2}_{-9.4}\right)~\mev.
\end{align}
From Fig.~\ref{fig:abst}, one sees clearly that the 1st pole on RS-I produces a peak of $t_{I=1}$ below the $pD^0$ threshold, while the RS-II pole would be partly responsible for the ``bump'' between the $pD^0$ and $nD^+$ thresholds. 
The two poles could be assigned to the $\Sigma_c(2800)^+$ resonance discovered by the Belle Collaboration~\cite{Mizuk:2004yu}, which has a mass of $2792^{+14}_{-5}$~MeV and a width of $62^{+60}_{-40}$~MeV.\footnote{Apart from the literature on the molecular interpretation of the $\Sigma_c(2800)$ cited in Sec.~\ref{sec:intro}, studies related to the $\Sigma_c(2800)$ at the quark level can be found in, e.g., Refs.~\cite{Copley:1979wj,Pirjol:1997nh,Roberts:2007ni,Garcilazo:2007eh,Ebert:2007nw,Zhong:2007gp,Chen:2007xf,Yoshida:2015tia,Chen:2015kpa,Shah:2016mig,Chen:2016iyi}. Concerning its spin and parity, no consensus has been reached so far.}
Because of the isovector nature, the $\Sigma_c(2800)$ signal is much weaker than those of excited $\Lambda_c$ states in the $\Lambda_b\to \pi^- p D^0$ decay. Yet, it leaves a footprint at the $nD^0$ threshold, producing a strong cusp. The threshold cusp effect can serve as a partial-wave filter since it shows up only for the $s$-wave.
The quantum numbers of the $\Sigma_c(2800)$ are not known yet, and our analysis suggests them to be $J^P=1/2^-$. This feature is shared by the results in a meson-exchange model in Ref.~\cite{Haidenbauer:2010ch}. Yet, we notice that a recent study of $D^{(*)}N$ interaction using chiral effective field theory and model inputs for the involved low-energy constants did not find a bound state in the isovector $ND$ system~\cite{Wang:2020dhf}. The analysis here should also be useful for improving the input of such studies.

In contrast, the second pole on RS-I couples dominantly to the isoscalar channel except when it is located on the real axis; furthermore, this pole in most of the parameter space is much further away from the thresholds, and one cannot make a unique conclusion on whether there is a near-threshold isoscalar pole (or even whether the isoscalar interaction is attractive or repulsive). Correspondingly, the curve for $|t_{I=0}|$ in most of the parameter space is quite smooth with two mild threshold cusps (as can be seen from the blue dashed best-fit curve in  Fig.~\ref{fig:abst}).

\section{Summary}
\label{sec:summary}

In this study, we have extracted the isoscalar and isovector $ND$ scattering lengths from the $pD^0$ invariant-mass distribution for the $\Lambda_b\rightarrow\pi^-pD^0$ decay reported by the LHCb Collaboration~\cite{Aaij:2017vbw}. In particular, there is a structure  in the data around the $nD^+$ threshold that is likely due to the $nD^+$ threshold cusp. 
The strength of the cusp is sensitive to the  $ND$ interaction, and thus can shed light on resonances near the $ND$ threshold.
Our analysis is based on a low-energy nonrelativistic effective field theory with $pD^0$ and $nD^+$ coupled channels. The effects of lower channels are built in by introducing imaginary parts to the scattering lengths.
From fitting to the $1/2^-$ partial wave of the $pD^0$ distribution, we find that the real parts of both the isoscalar and isovector $ND$ scattering lengths are negative, implying that the interactions are either repulsive or have a bound state below threshold. 
Within uncertainties, the real part of the isoscalar $ND$ scattering length ranges from $-1.4$ to $-0.1$~fm, and thus one is not able to conclude whether there must be a near-threshold isoscalar $\Lambda_c$ excited state or the interaction is repulsive.
In contrast, the absolute value of the real part of the isovector $ND$ scattering length is always large, indicating the existence of a bound-state pole below the $pD^0$ threshold. 
We indeed find a bound state at $2801.8^{+1.0}_{-4.0} - i\,(2.6\pm2.6)~\mev$; another pole in RS-II is found at $2807.0^{+4.5}_{-17.1} - i\left(9.4^{+10.2}_{-9.4}\right)~\mev$. Both of them couple dominantly to the isovector channel, and in the isospin limit only one bound-state pole is left, suggesting they correspond to a two-pole structure of the same state due to coupled channels.
This state could be assigned to the $\Sigma_c(2800)$ discovered by the Belle Collaboration~\cite{Mizuk:2004yu}, and its quantum numbers are suggested to be $J^P=1/2^-$.
Data with better statistics would be helpful to further clarify this situation.

\section*{Acknowledgements}
F.-K.G.\ is grateful to Anton Poluektov for helpful communications. This work is supported in part by the National Natural Science Foundation of China (NSFC) and  the Deutsche Forschungsgemeinschaft (DFG) through the funds provided to the Sino--German Collaborative Research Center  CRC110 ``Symmetries and the Emergence of Structure in QCD"  (NSFC Grant No.~11621131001), by the NSFC under Grants No.~11835015, No.~11947302, and No.~11961141012, by the Chinese Academy of Sciences (CAS) under Grants No.~XDB34030303 and No.~QYZDB-SSW-SYS013, and by the CAS Center for Excellence in Particle Physics (CCEPP).
S.S.\ is also supported by the 2019 International Postdoctoral Exchange Program, and by the CAS President's International Fellowship Initiative (PIFI) under Grant No.~2019PM0108.

\bigskip

\bibliographystyle{elsarticle-num}
\interlinepenalty=10000
\bibliography{refs}

\end{document}